# SAERMA: Stacked Autoencoder Rule Mining Algorithm for the Interpretation of Epistatic Interactions in GWAS for Extreme Obesity

Casimiro A. Curbelo Montañez, Paul Fergus, Carl Chalmers, Nurul H. A. Hassain Malim, Basma Abdulaimma, Denis Reilly, and Francesco Falciani

**Abstract**—One of the most important challenges in the analysis of high-throughput genetic data is the development of efficient computational methods to identify statistically significant Single Nucleotide Polymorphisms (SNPs). Genome-wide association studies (GWAS) use single-locus analysis where each SNP is independently tested for association with phenotypes. The limitation with this approach, however, is its inability to explain genetic variation in complex diseases. Alternative approaches are required to model the intricate relationships between SNPs. Our proposed approach extends GWAS by combining deep learning stacked autoencoders (SAEs) and association rule mining (ARM) to identify epistatic interactions between SNPs. Following traditional GWAS quality control and association analysis, the most significant SNPs are selected and used in the subsequent analysis to investigate epistasis. SAERMA controls the classification results produced in the final fully connected multi-layer feedforward artificial neural network (MLP) by manipulating the interestingness measures, support and confidence, in the rule generation process. The best classification results were achieved with 204 SNPs compressed to 100 units (77% AUC, 77% SE, 68% SP, 53% Gini, logloss=0.58, and MSE=0.20), although it was possible to achieve 73% AUC (77% SE, 63% SP, 45% Gini, logloss=0.62, and MSE=0.21) with 50 hidden units – both supported by close model interpretation.

---

## 1 INTRODUCTION

UNDERSTANDING the genetic architecture of common diseases remains a significant challenge. Advances in the field have identified genetic variations that underlie common disorders such as obesity, type 2 diabetes, and certain cancers [1]. However, we are no closer to identifying the precise genetic markers that result in the manifestation of complex phenotypes.

Single nucleotide polymorphisms (SNPs) [2] are the most common type of genetic variation among humans. These have become the genetic marker of choice in the genetic mapping of complex traits. Genome-Wide Association Studies (GWAS) [3] utilise SNP Information and this has helped to improve our knowledge and understanding of disease genetics. GWAS implements single-loci analysis where SNPs are independently tested for association with phenotypes of interest, without consideration of the interactions that occur between loci. This is a major limitation in GWAS, particularly when studying complex disorders

- C. A. Curbelo Montanez, P. Fergus, C. Chalmers, Basma T. Abdulaimma, and Denis Reilly are from the Faculty of Engineering and Technology, Liverpool John Moores University, Liverpool, UK. E-mail: c.a.curbelomontanez@ljmu.ac.uk, p.fergus@ljmu.ac.uk, c.chalmers@ljmu.ac.uk, b.t.abdulaimma@2015.ljmu.ac.uk, and d.reilly@ljmu.ac.uk.
- N. H. A. H. Malim is with the School of Computer Sciences, Universiti Sains Malaysia, 11800, Pulau Pinang, Malaysia. E-mail: nurulhashimah@usm.my.
- Francesco Falciani is with the department of Functional and Comparative Genomics, Institute of Integrative Biology, Liverpool university, Liverpool, UK. E-mail: f.falciani@liverpool.ac.uk

caused by SNP-SNP, gene-gene and gene-environment interactions. Therefore, to better understand the missing heritability inherent in GWAS it is necessary to examine epistasis interactions [4]. This approach assumes that genes do not work independently but create "gene networks" that have major effects on tested phenotypes. Hence, identifying epistatic interactions will help us to understand biological mechanisms and predict complex traits from genotype data.

Combinatorial effects between genes are termed epistatic interactions or epistasis [5]. Different perspectives exist: biological (or functional epistasis) and statistical epistasis [5]. Statistical epistasis is investigated in this paper as it provides a suitable strategy for discovering new gentic pathways. This provides a foundation for new discoveries and testable hypotheses.

Traditional statistical methods such as logistic regression have shown limited power in modelling high-order nonlinear interactions between genetic variants [6]. Additionally, the high dimensionality present in genomic data makes it computationally difficult to exhaustively evaluate all SNP combinations. This is a well-known computational challenge in the field of computer science [7]. Finally, it is important to interpret gene-to-gene interactions in the context of human biology before any results can be translated into specific recommendations and treatment strategies. However, making etiological inferences from computational models has been considered the most relevant but difficult challenge [8].

One approach worth considering is Association Rule Mining (ARM), which is an unsupervised learning method used to find relationships between items (variables) that co-occur in large data sets [9]. The discovery of association rules depends on the discovery of frequent itemsets, where association rules are required to satisfy *support* and *confidence* user defined constraints. This technique has been used to discover binding cores in protein-DNA [10] and to find associations between the regulation of gene expression levels and phenotypic variations in gene expression analysis [11]. An application of the Apriori algorithm [12], in the context of case-control association studies and epistasis analysis, is AprioriGWAS [13]. This tool was applied to age-related macular degeneration (AMD) and bipolar disorder (BD) data with promising interactions between genes found. The approach in [13] uses frequent itemset mining (FIM) with Apriori to look for genotype patterns with different frequencies in cases and controls.

With regards to the discovery of SNP-to-SNP interactions, deep learning (DL) has shown promise. Deep learning is a type of ANN and one of the most active fields in machine learning today. DL architectures have proved to be particularly useful in image and speech recognition, natural language understanding and most recently, in computational biology [14]. They are characterised by deep hidden layers and neurons. In Bioinformatics, DL has been used to select regulatory SNPs with functional impact before association analysis is conducted (DeepWAS) [15]. The study focused on variants (SNPs) that alter functional regulatory elements (i.e. elements that control gene expression and DNA methylation) which are identified using a deep learning-based algorithmic framework: DeepSEA [16].

This paper extends these works and combines ARM and DL techniques to investigate genetic epistasis in obesity. Obesity is considered one of the most difficult clinical and health challenges worldwide [17]. It has become a global epidemic, also contributing to the growing rates of type 2 diabetes (T2D) and cardio vascular disease among other non-communicable diseases [18]. The ubiquitous availability of low-cost hypercaloric food combined with a sedentary lifestyle and other environmental factors, have played a fundamental role in the obesity epidemic. Surprisingly, not every individual exposed to such environments, also known as obesogenic environments, becomes obese. Therefore, the lack of understanding about the mechanisms that underlie individual differences in the predisposition to obesity have motivated this study. While GWAS has identified several variants associated with obesity traits (i.e. FTO and MC4R), they do not explain the variability of obesity attributable to genetic factors. Interactions between genes, namely epistasis, will help to provide a better understanding of polygenic obesity. This is regarded as a much more intuitive approach given that complex diseases cannot be reduced to single univariate SNP-phenotype interactions.

In body mass index (BMI) and obesity GWAS, gene-gene interactions have received little attention [19]. Thus, a novel methodology is considered in this paper, in which a subset of loci after quality-control (QC) and association analysis was selected (statistical filtering). Epistatic interactions within the remaining genetic variants are investigated using deep learning stacked autoencoders (SAE) and ARM. Basic statistical analysis methods and techniques for the analysis of genetic SNP data from population-based genome-wide studies are considered, particularly logistic regression. Subsequent analysis of epistasis is carried out using SAEs to learn the deep features and, ARM with the Apriori algorithm, to discover a set of frequent patterns expressed as association rules. Both, SAE and ARM describe relationships between SNPs in extreme cases of obesity (Body Mass Index (BMI) > 40 kg/m$^2$) and normal samples from a subset of cases and controls within the Geisinger MyCode project [20]. The performance of the features selected by ARM and those extracted by SAE are objectively measured using a multi-layer feedforward artificial neural network (MLP).

## 2 MATERIALS AND METHODS

This section introduces the data used in the study, quality control (QC), and association and statistical epistasis analysis.

### 2.1 eMERGE Data

Case-control data was obtained from the database of Genotypes and Phenotypes (dbGaP) [21]. Controls were obtained from the eMERGE Geisinger eGenomic Medicine (GeM) - MyCode Project Controls (dbGaP study accession phs000381.v1.p1), while cases were obtained from the eMERGE Genome-Wide Association Studies of Obesity project (dbGaP study accession phs000408.v1.p1).

The case-control dataset contains 2,193 participants (917 males and 1,236 females). Each participant contains 594,034 genetic markers. Furthermore, 99.5% of the participants are from a white ethnic background (Caucasians).

### 2.2 Data Pre-processing

Individuals reported as white were selected to conduct GWAS to reduce potential bias caused by population stratification [22]. QC and filtering procedures were performed on individuals and SNPs, following standard QC protocols and guidance in [22].

Samples with discordant sex were removed. Related and duplicate samples were removed using Identity by Descent (IBD) coefficient estimates (IBD > 0.185). Principal component analysis (PCA) was performed to identify outliers and hidden population structure using EIGENSTRAT [23]. SNPs with minor allele frequency (MAF) lower than 5%, Hardy-Weinberg Equilibrium (HWE) P-value lower than $1 \times 10^{-5}$ in control subjects and, a call rate lower than 99% were excluded from further analysis. After QC, 1,997 individuals (879 cases and 1118 controls) and 240,950 genetic variants were retained for subsequent analysis.

### 2.3 Statistical Filtering

Association analysis is used to reduce the computationally large number of genetic variants (240,950 SNPs). Statistical association testing between SNPs and the obesity phenotype was conducted under an additive model using logistic regression [24].

Statistical filtering is used to reduce SNPs with insignificant marginal effects to meet the computational needs required for epistatic analysis and machine learning tasks. Therefore, only SNPs with P-values lower than $1\times10^{-2}$ are utilised for detecting epistatic interactions and to minimise computational overheads.

## 2.4 Multi-layer Feedforward Artificial Neural Network

A multi-layer feedforward artificial neural network (ANN) is implemented based on the formal definitions in [25], to conduct binary classification. A feedforward ANN (FNN) is utilised with labelled training samples $(x^{(i)}, y^{(i)})$ from case-control genetic data where $y^{(i)} \in \mathbb{R}^2$, to train the network for supervised learning tasks. A non-linear hypothesis $h_{W,b}(x)$ is defined using FNN, with parameters $W,b$ fitted to the data. The parameter $x$ is a vector of input features representing individuals while outputs for the two class labels (obese or non-obese) are represented using $y$. The weight and bias parameters are learnt by minimising the cost function with stochastic gradient descent [26]. The learning process is performed using the back-propagation algorithm and gradient descent [27].

## 2.5 Autoencoders

Deep feedforward Autoencoders (AE) are used in this study for unsupervised feature learning and non-linear dimensionality reduction [28]. An AE is a three-layer neural network that learns an output $\hat{x}$ that is similar to the input $x$. Hence, an AE tries to learn a function $h_{W,b}(x) \approx x$, given a set of unlabeled training samples $\{x^{(1)}, x^{(2)}, x^{(3)}, ...\}$, where $x^{(i)} \in \mathbb{R}^n$. The second layer or hidden layer generates the deep features by minimizing the error between the input vector and the reconstructed ouput vector.

First, the encode phase maps input data into a feature vector $z$ so that, for each sample $x^{(i)}$ from the input set $\{x^{(1)}, x^{(2)}, x^{(3)}, ...\}$, we have

$$z^{(i)} = f(W^{(1)}x^{(i)} + b^{(1)}) \qquad (1)$$

while in the decode phase, the decoder reconstructs the input $x$, producing a reconstructed space $\hat{x}$ defined as

$$\hat{x}^{(i)} = f(W^{(2)}z^{(i)} + b^{(2)}) \qquad (2)$$

where $W^{(1)}$ and $W^{(2)}$ represent the input-to-hidden and the hidden-to-output weights respectively, $b^{(1)}$ and $b^{(2)}$ represent the bias of hidden and output neurons, whereas $f(\cdot)$ denotes the activation function.

Parameters $W^{(1)}$, $W^{(2)}$, $b^{(1)}$ and $b^{(2)}$ in the AE are learnt by minimising the reconstruction error

$$J(W,b;x,\hat{x}) = \frac{1}{2}\left\| h_{W,b}(x) - \hat{x} \right\|^2. \qquad (3)$$

This is a measurement of discrepancy between input $x$ and reconstructed $\hat{x}$ with respect to a single sample. For a training set of $m$ samples, the cost function of an autoencoder is defined as:

$$J(W,b) = \left[ \frac{1}{m}\sum_{i=1}^{m} J\left(W,b;x^{(i)},\hat{x}^{(i)}\right) \right] + \frac{\lambda}{2}\sum_{l=1}^{n_l-1}\sum_{i=1}^{s_l}\sum_{j=1}^{s_l+1}\left(W_{ji}^{(l)}\right)^2$$

$$= \left[ \frac{1}{m}\sum_{i=1}^{m}\left(\frac{1}{2}\left\| h_{W,b}\left(x^{(i)}\right) - \hat{x}^{(i)} \right\|^2\right) \right] + \frac{\lambda}{2}\sum_{l=1}^{n_l-1}\sum_{i=1}^{s_l}\sum_{j=1}^{s_l+1}\left(W_{ji}^{(l)}\right)^2$$

$$(4)$$

where $m$ denotes the overall training set size, $s$ denotes the number of nodes in layer $l$, $\lambda$ is the weight decay parameter and the square error is used as the reconstruction error for each training sample. The second term is introduced to decrease the magnitude of the weights which helps prevent overfitting. Equation (4) can be minimised using stochastic gradient descent.

AEs are stacked layer by layer to produce a Stacked Autoencoder (SAE) [29]. Once a single layer AE has been trained, a second AE is trained using the hidden layer from the first AE as shown in Fig. 1. By repeating this procedure, it is possible to create SAEs of arbitrary depth.

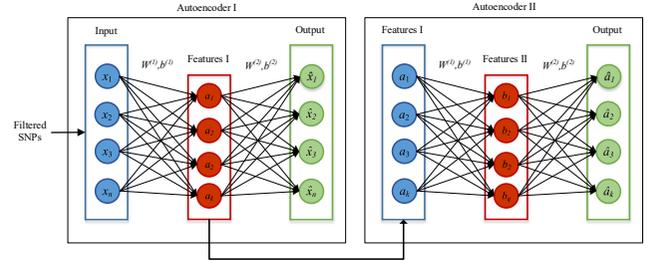

Fig. 1. Example of SAE formed by two single AEs

AEs are stacked to enable greedy layer-wise learning where the $l_{th}$ hidden layer is used as input to the $l+1$ hidden layer in the stack. The results produced by the SAE are utilised to pre-train the weights of a fully connected MLP, rather than randomly initialising the weights to small values. This approach helps models initialise parameters near to a good local minimum and improve optimisation. This shows that smoother convergence and higher overall performance in classification tasks is possible using this approach.

An SAE with 2,000, 1,000, 500 and 50 hidden neurons in each hidden layer is used during experimentation. The proposed SAE architecture extracts a mapping that decodes the input (set of SNPs) as closely as possible without losing significant SNP-SNP patterns.

The SAE configuration decreases the dimensionality of the original data stack by stack, leading to a reduction in noise while preserving the most important information for MLP tuning. The complexity of this approach is that it is difficult to determine what SNPs contribute to classfication accuracy. ANN models in general are very difficult to interpret. Therefore, SAEs are combined with assocaition rule mining to describe what SNPs and associated interactions contribute to classification results.

## 2.6 Association Rule Mining (ARM)

Association rules are implemented to reveal biologically relevant associations between SNPs. If SNPs frequently appear together, there is an underlying relationship between them. Exploring the intrinsic relationships in the data is performed using frequent pattern mining (FPM). This technique extracts all frequent itemsets from a dataset, which are then used to generate association rules. In the proposed method, the idea is to extract important rules identified in cases and controls separately.

Using association rule mining, frequently occurring SNPs as items are identified (itemsets) in different individuals as transactions. In other words, individuals are transactions, SNPs are items, and SNP combinations are itemsets. Single SNPs tend to have small effect sizes in polygenic diseases. Therefore, by looking at the joint effect of multiple SNPs, explanatory power can be increased.

Typically, ARM assumes a common strategy for decomposing mining problems into two principal subtasks: 1) *Frequent itemset generation* and, 2) *rule generation*.

Itemsets are sets of $k$-items where $k$ starts with 1 to infinity. Unnecessary itemset candidates are produced if at least one of its subsets is infrequent. Hence, the frequent itemset generation is equipped with pruning steps to eliminate $k$-itemset candidates based on a minimum support threshold. Support is the number of transactions that contain a particular itemset.

Frequent itemsets are independent sets of SNPs (itemsets) in the Geisinger MyCode dataset whose support is greater than or equal to a given minimum support threshold $\sigma$. Itemsets whose support count is lower than the minimum $\sigma$, are removed. This strategy based on support measures is termed support-based pruning.

Once frequent itemsets have been obtained the generation of association rules is performed. Association rule mining discovers sets of SNPs that frequently occur together in the MyCode dataset and creates a relationship between those SNPs in the form $X \rightarrow Y$. This relationship implies that when $X$ occurs it is likely that $Y$ also occurs. Such a relationship is called an association rule. An association rule is defined as an implication of the form $X \rightarrow Y$, where $X, Y \subseteq I$ and $X \cap Y = \emptyset$. $X$ refers to the left-hand side (LHS) or antecedent of the rule, $Y$ is the right-hand side (RHS) or consequent, and $I$ is a set of items.

Given a set of transactions $T$, ARM searches for all the rules with *support* $\geq \sigma$ and *confidence* $\geq \delta$ where $\sigma$ and $\delta$ are the corresponding minimum support and confidence thresholds. Support and confidence are formally defined as (5) and (6) respectively.

$$support(X \rightarrow Y) = \frac{support(X \cup Y)}{|T|} \qquad (5)$$

$$confidence(X \rightarrow Y) = \frac{support(X \cup Y)}{support(X)}. \qquad (6)$$

Rules are generated from each of the frequent $k$-itemsets. Hence, the total candidate association rules produced can be up to $2^k - 2$, excluding those that are null in the antecedent ($X$) or consequent ($Y$).

The significance of the association rules is measured in terms of their *support* and *confidence* although other interest measures such as *lift* or *Chi-Square* can be used to validate rules. The support of a rule is the probability that the samples in a dataset contain both $X$ and $Y$. Rules with very low support may occur by chance, therefore, support is an important measure that can be used to eliminate unimportant rules. Confidence of a rule, on the other hand, is the probability that a case contains $Y$ given that it contains $X$. It provides an estimate of the conditional probability of $Y$ given $X$, $P(Y|X)$.

### 2.6.1 *Apriori algorithm*

The generation of association rules is conducted using the Apriori algorithm [12]. The Apriori algorithm performs a breadth-first search (BFS), enumerating every single frequent itemset by iteratively generating candidate itemsets. Candidate itemsets of length $k$ are generated from $k-1$ itemsets. The support of every candidate itemset is calculated iteratively where itemsets with support values under a defined threshold are disregarded.

To manage the very large number of discovered association rules, the patterns are filtered, grouped and organized. This is a crucial step to focus on the most interesting association rules. Nearly all search algorithms rely on support-based pruning. If an itemset $X$ is not frequent (given a minimum support), then none of its supersets $Y \supset X$ can be frequent. This property is known as anti-monocity of the frequency. Furthermore, if the support value is set too low (close to 0), a large number of spurious rules are generated. This makes the problem computationally intractable. Conversely, if the value for support is too high (close to 1), a very small number of rules (or no rules at all) are extracted, which means several significant rules could be missed. Accepting or rejecting spurious patterns (rules) is known in statistics as type 1 and type 2 errors respectively. To reduce type errors, the traditional support-confidence framework is replaced by a support-dependence framework.

Standard minimum support and confidence measures set by the user are employed by the algorithm to prune uninterested association rules. However, the minimum frequency and confidence requirements do not guarantee statistical dependence or significance. Hence, it is also possible to add additional objective interest measures to each rule, e.g. the P-value threshold computed using the Chi-square test or Fisher's exact test to evaluate the significance of the rules.

### 2.6.2 *Additional Interest Measures*

Limitations with support-confidence based rule mining [30] has resulted in other interestingness measures to evaluate the quality of the patterns identified. Examples of these measures are lift, P-value thresholds computed using the Chi-square test or Fisher's Exact test, and a collection of other objective symmetric and asymmetric interestingness methods [30]. In this study, those described previously are used in addition to lift and Chi-squared to determine significant rules, as they allow us to measure which rules are more correlated.

Lift or interest, is a symmetric measure which divides the rule's confidence by the support of the itemset in the rule consequent as shown in (7). It can be used to analyse the relativity of association rules mined and for measuring how many times more frequently $X$ and $Y$ occur together than expected under statistically independent conditions. Lift indicates a positive correlation between $X$ and $Y$ when its value is greater than one, negative correlation when its value is lower than one, and independence when lift is equal to one. As an example, a lift($X \to Y$) > 1 indicates that the appearance of $X$ promotes the appearance of $Y$, resulting in a positive correlated rule. Thus, the higher the lift, the stronger the positive correlation and the more dependent the SNPs are. In this paper, only positive correlated rules are of interest

$$lift(X \to Y) = \frac{confidence(X \to Y)}{support(Y)} = \frac{support(X \cup Y)}{support(X)\,support(Y)}. \quad (7)$$

Finding measures that can be used with lift to make the best selection of rules is crucial. Despite the numerous alternatives for expressing the dependence between the antecedent and the consequent of an association rule, the classic Chi-square test statistic ($\chi^2$), can be used to determine the statistical significance level of association rules [31]. Thus, rules can be pruned in case of independency, meaning that the itemsets (SNPs) in the rule are not correlated. $\chi^2$ helps deciding whether items in the rules are independent of each other, but it is not useful for ranking purposes by itself. The standard Chi-squared test statistic ($\chi^2$) is defined as:

$$\chi^2 = \sum \frac{(f_0 - f)^2}{f}. \quad (8)$$

$\chi^2$ is a summed normalized square deviation of the observed values from the expected values. An important fact about the Chi-square test is that it can be used to calculate the P-value to determine the significance level of the rule. For instance, if the P-value of the rule is lower than 0.05, that is a $\chi^2$ value higher than 3.84, we can tell that $X$ and $Y$ are significantly dependent and, therefore, the rule $X \to Y$ can be considered for subsequent analysis. This is one way to identify the direction of a rule when summarizing unpruned rules, by the type of correlation the rules have, as similarly performed by lift (positive correlation, negative correlation or independence). To some extent, $\chi^2$ improves the traditional framework of the interestingness measure provided by lift.

A combination of different interest measures is necessary to assess the strength and the dependency of the antecedent and consequent of the rules. Discovered associations are pruned to remove non-significant rules, and then a special subset of the unpruned associations forms a summary of the discovered associations which represent candidates for epistatic interactions.

### 2.6.3 Redundancy

Redundancy elimination tasks can be beneficial to reduce complexity by identifying smaller sets of more general rules which are easier to interpret than larger complex, and frequently overlapping rules. Rules are considered redundant, if a more general rule or rules with the same or higher confidence values are present. Formally, for $X'$ subset of $X$, a rule $X \to Y$ is redundant if,

$$confidence(X' \to Y) \geq confidence(X \to Y). \quad (9)$$

The idea is to find statistically significant rules after support and confidence pruning, in addition to redundant rule elimination. For this reason, several assumptions have been considered to rank the rules. First, the rules must be common in, at least, 60% of the individuals. Second, the higher the confidence the more likely it is for $Y$ to be present in transactions that contain $X$. According to this, a support value of 0.6 and a confidence value of 0.8 are used to generate rules in this study.

### 2.7 Model Performance

The exactness of a classification can be evaluated by computing a contingency table. In this study, classifier performance is assessed through sensitivity (SE), specificity (SP), gini, logloss, area under the curve (AUC) and mean squared error (MSE) as performed in [32], [33]. Classifiers with good predictive capacity possess SE, SP, gini and AUC values close to 1 with logloss and MSE values close to 0. Additionally, hyperparameter optimisation is performed using random search [34]. Random search has proven to be as good as, or even better than, pure grid search when applied to ANNs, saving computational time [34]. This is true since random grid search can effectively search a larger, and often less promising, configuration space.

## 3 Results

In this section, the results are presenteed using the proposed methodology outlined above. This is reported in four experiments conducted after QC and association analysis (Statistical filtering): 1) Baseline classification with GLM using SNPs with P-value < $10^{-2}$; 2) MLP classification using SNPs with P-value < $10^{-2}$; 3) SAE-based classification using non-linear SNP-SNP interactions with P-values < $10^{-2}$; and 4) our proposed approach, SAERMA .

Genotype data for 960 cases and 1,223 control subjects were analysed. After QC, a total of 240,950 variants and 1,997 individuals passed subsequent filter analysis and QC. Among the remaining phenotypes, 879 are cases and 1,118 are controls. These are used to conduct association analysis of extreme obesity trait as a statistical filtering approach. The results from association tests with P-value < $1 \times 10^{-2}$ are considered, resulting in a subset of 2,465 SNPs. The resulting outcomes are therefore, considered as hypothesis generating.

### 3.1 GLM Classification

The first experiment conducted following QC and association analysis utilises classification tasks and the filtered SNPs (2,465). Before conducting experiments with more complex approaches such as ANNs or SAEs, classification is performed with a generalised linear model (GLM) [35].

Four different sets of SNPs (5, 32, 248 and 2,465 SNPs) were derived using different $P$-value thresholds as indicated in Table 1, and used to train a GLM to classify extremely obese and non-obese observations. The data set is split randomly into training (60%), validation (20%) and testing (20%).

**TABLE 1**
**SET OF SNPs SELECTED**

| Set | P-value | Number of SNPs |
|-----|---------|----------------|
| 1 | $1\times10^{-5}$ | 5 |
| 2 | $1\times10^{-4}$ | 32 |
| 3 | $1\times10^{-3}$ | 248 |
| 4 | $1\times10^{-2}$ | 2,465 |

Regularisation parameters *alpha* and *lambda* were tuned, and the optimal values were obtained using a random search. Based on empirical analysis, *alpha* and *lambda* values for set 1 (alpha=0.5 and lambda=0.00598), set 2 (alpha=0.5 and lambda=0.00204), set 3 (alpha=0.5 and lambda=0.00970) and set 4 (alpha=0.5 and lambda=0.00151) produced the best classification results.

Using optimised F1 threshold values 0.3527, 0.4532, 0.3969 and 0.6684 the results in the validation set were obtained as shown in Table 2 for 5 SNPs ($1\times10^{-5}$), 32 SNPs ($1\times10^{-4}$), 248 SNPs ($1\times10^{-5}$) and 2,465 SNPs ($1\times10^{-2}$) respectively.

**TABLE 2**
**PERFORMANCE METRICS FOR VALIDATION SET**

| Set | SE | SP | Gini | LogLoss | AUC | MSE |
|-----|-----|-----|------|---------|-----|-----|
| 1 | 0.8723 | 0.2819 | 0.2563 | 0.6619 | 0.6281 | 0.2348 |
| 2 | 0.6862 | 0.7225 | 0.5010 | 0.5865 | 0.7505 | 0.2004 |
| 3 | 0.8298 | 0.8194 | 0.8081 | 0.3938 | 0.9041 | 0.1261 |
| 4 | 0.7606 | 0.9383 | 0.8317 | 0.3841 | 0.9158 | 0.1150 |

The performance metrics for the test set are shown in Table 3. These metric values were obtained using optimised F1 thresholds 0.2893, 0.4533, 0.2368 and 0.4665 for $1\times10^{-5}$, $1\times10^{-4}$, $1\times10^{-3}$ and $1\times10^{-2}$, respectively.

**TABLE 3**
**PERFORMANCE METRICS FOR TEST SET**

| Set | SE | SP | Gini | LogLoss | AUC | MSE |
|-----|-----|-----|------|---------|-----|-----|
| 1 | 0.9774 | 0.0909 | 0.2145 | 0.6736 | 0.6073 | 0.2404 |
| 2 | 0.9548 | 0.2440 | 0.4186 | 0.6185 | 0.7093 | 0.2153 |
| 3 | 0.9548 | 0.6316 | 0.7798 | 0.4119 | 0.8899 | 0.1350 |
| 4 | 0.8531 | 0.9043 | 0.8725 | 0.3288 | 0.9362 | 0.0976 |

The ROC curve comparison depicted in Fig. 2 is used as a graphical performance measure to summarise the predictive performance of the GLM models. The cut-off values for the false and true positive rates using the test set are shown in each of the ROC curves for the different implemented classifiers. In this first evaluation, there is a clear deterioration in performance as the number of SNPs decreases (*P*-value threshold increases). Note that SNPs with conservative *P*-value thresholds are an indication of how significant associations are. This demonstrates the limitations of the most significant SNPs in classifying case-control samples. The highest performance was obtained with 2,465 SNPs and the lowest with 5 SNPs.

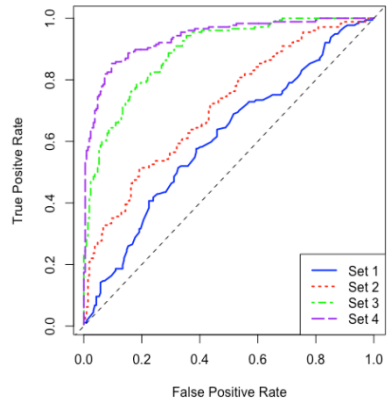

Fig. 2: ROC curves for the test set using GLM models trained with different P-value thresholds

## 3.2 MLP Classification

Using the statistical filtering results in Table 1, an MLP is trained and used for classification analysis using the same performance metrics. For each MLP model, the network architecture and associated regularization parameters were tuned. This was achieved using random search and a maximum of 200 models. Model training is stopped when the logloss value fails to improve by at least 1% (stopping tolerance) for two scoring events (stopping rounds). The adaptive learning rate ADADELTA [36] was used for stochastic gradient descent optimisation, with parameters *rho* and *epsilon* set to 0.99 and $1\times10^{-8}$ respectively, to balance the global and local search efficiencies.

To prevent overfitting, stability and improved generalisation, Lasso ($L_1$) and Ridge ($L_2$) regularisation, and input dropout ratio are all tuned. $L_1$ only allows strong weights to survive, $L_2$ prevents them from getting too big and input dropout ratio regulates the number of neurons randomly dropped in the input layer - hidden dropout ratios do the same in hidden layers. Based on empirical analysis, these configurations produced the best results.

The performance metrics for the validation set are provided in Table 4. The results show the four SNP configurations in Table 1, using optimized F1 threshold values 0.2674, 0.4463, 0.3551 and 0.8084, respectively.

Table 5 shows the performance metrics for the test data using optimised F1 thresholds 0.2675, 0.2157, 0.4312 and 0.6303 for $1\times10^{-5}$, $1\times10^{-4}$, $1\times10^{-3}$ and $1\times10^{-2}$, respectively. The results are generally lower than those achieved with the validation set but, in some cases, not by much.

**TABLE 4**
**PERFORMANCE METRICS FOR VALIDATION SET**

| Set | SE | SP | Gini | LogLoss | AUC | MSE |
|-----|-----|-----|------|---------|-----|-----|
| 1 | 0.9415 | 0.1806 | 0.2556 | 0.6606 | 0.6278 | 0.2342 |
| 2 | 0.6915 | 0.7490 | 0.5117 | 0.5828 | 0.7558 | 0.1987 |
| 3 | 0.8564 | 0.8194 | 0.8474 | 0.3510 | 0.9237 | 0.1120 |
| 4 | 0.9628 | 0.9780 | 0.9923 | 0.0883 | 0.9961 | 0.0259 |

TABLE 5
PERFORMANCE METRICS FOR TEST SET

| Set | SE | SP | Gini | LogLoss | AUC | MSE |
|-----|-----|-----|-----|---------|-----|-----|
| 1 | 0.9943 | 0.0622 | 0.2074 | 0.6750 | 0.6037 | 0.2410 |
| 2 | 0.9491 | 0.2871 | 0.4331 | 0.6151 | 0.7165 | 0.2140 |
| 3 | 0.9039 | 0.7942 | 0.8512 | 0.3476 | 0.9256 | 0.1094 |
| 4 | 0.9548 | 0.9761 | 0.9878 | 0.1061 | 0.9938 | 0.0291 |

TABLE 6
PERFORMANCE METRICS FOR VALIDATION SET

| Layers | SE | SP | Gini | LogLoss | AUC | MSE |
|--------|-----|-----|-----|---------|-----|-----|
| 1 AE | 0.9202 | 0.9383 | 0.9608 | 0.1817 | 0.9804 | 0.0547 |
| 2 AEs | 0.8404 | 0.9383 | 0.9034 | 0.2889 | 0.9517 | 0.0848 |
| 3 AEs | 0.8670 | 0.8899 | 0.8828 | 0.3146 | 0.9414 | 0.0963 |
| 4 AEs | 0.9202 | 0.5771 | 0.6976 | 0.4776 | 0.8488 | 0.1593 |

AE = Auto Encoder

The ROC curves in Fig. 3 illustrate the cut-off values for the false and true positive rates using the test set. There is a clear deterioration in performance as the number of SNPs decreases (P-value threshold increases). In this instance, the results highlight the limited predictive capacity of highly ranked SNPs when discriminating between case and control samples.

Table 7 shows the performance metrics obtained using the test set. Optimised F1 threshold values 0.5363, 0.3356, 0.3899 and 0.4615 were used to obtain these metrics with models trained on 2,000, 1,000, 500 and 50 compressed input units respectively.

TABLE 7
PERFORMANCE METRICS FOR TEST SET

| Layers | SE | SP | Gini | LogLoss | AUC | MSE |
|--------|-----|-----|-----|---------|-----|-----|
| 1 AE | 0.9491 | 0.9330 | 0.9499 | 0.1956 | 0.9750 | 0.0540 |
| 2 AEs | 0.9152 | 0.8756 | 0.9102 | 0.2948 | 0.9551 | 0.0875 |
| 3 AEs | 0.9096 | 0.8756 | 0.9005 | 0.2851 | 0.9502 | 0.0872 |
| 4 AEs | 0.7853 | 0.7990 | 0.7036 | 0.4769 | 0.8518 | 0.1563 |

AE = Auto Encoder

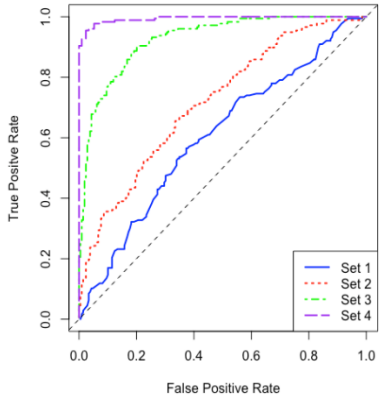

Fig. 3: ROC curves for test set using the MLP trained with different P-value thresholds

The cut-off values for false and true positive rates in the test set are depicted in Fig. 4. The ROC curves show a gradual deterioration in classifier performance as the initial 2,465 features (SNPs) are progressively compressed to 50 hidden units in the SAE. Despite the observable deterioration, the results remain high with 50 compressed hidden units. This is in stark contrast to the P-value approach adopted in the previous experiments with GLM and MLP without SAE weight initialisation.

### 3.3 Epistatic interactions using Stacked Autoencoders

In this evaluation, a SAE configuration is utilised to learn the deep features that exist in a subset of 2,465 SNPs (P-value $< 1 \times 10^{-2}$), to capture information about important SNPs and the cumulative epistatic interactions between them. A layer wise approach is adopted by stacking four single layer AEs with 2,000-1,000-500-50 hidden units, where the original 2,465 SNPs are compressed into progressively smaller hidden layers. The final SAE hidden layer is used to initialise the weights of an MLP. The data set is again randomly split into training (60%), validation (20%) and testing (20%), while hyperparameter tuning is performed through random search.

To measure the performance, each MLP classifier was initialized using different compressed unit configurations obtained from the SAE. Performance metrics for the validation and test sets are provided in Table 6 and Table 7 respectively. Using 2,000 hidden units, an optimised F1 threshold value of 0.4977 is assigned to extract the validation set metrics as indicated in Table 6. Successive layers of the SAE are used to initialise and fine-tune the remaining MLP models with 1,000, 500 and 50 hidden units respectively and F1 threshold values 0.6188, 0.4978 and 0.2701 for each of the remaining models respectively.

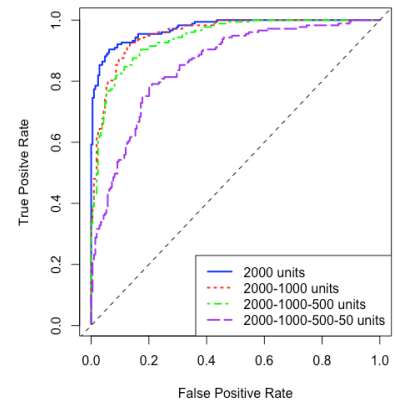

Fig. 4. ROC curves for the test set using trained models with the different compressed units considered for the SAE

### 3.4 SAERMA: Stacked Autoencoder Rule Mining Algorithm

In the final experiment QC, association analysis, rule mining, SAE and MLP classification are combined to form the SAERMA algorithm. Classification analysis in this experiment was conducted using a second feature selection step based on ARM. Rule mining allows us to find the most frequent SNPs (from the 2,465 SNPs considered) among individuals in cases and controls and then extract rules from them. The top 10 rules identified using the Apriori algorithm in cases and controls are listed in Table 8 and Table 9. As shown in Supplemental Material, Figure S1 in File 1, these rules can be plotted to provide insights through rule inspection.

Items from the rules (SNPs) are utilised as input features in our SAE for deep feature extraction (which includes the relationships between SNPs) and to initialise the weights of an MLP before fine-tuning for classification analysis. By adjusting support and confidence parameters in the rule generation process, the number of rules can be increased or decreased. This, in turn, impacts the performance of the SAE-MLP models generated for feature extraction and classification tasks. The results in this section are, therefore, derived from the SNPs contained within the most significant rules extracted with support $\sigma$ = 0.6 and confidence $\delta$ = 0.8 as discussed in this paper. These are the lowest interest measure values which allow rule generation without overloading the system used in this study.

#### TABLE 8
#### TOP 10 RULES IDENTIFIED IN CASES

| # | Rule | $\sigma$ | $\delta$ | Lift | $\chi^2$ |
|---|------|----------|----------|------|----------|
| 1 | [rs12053340_C_D] => [rs1527944_T_D] | 0.61 | 1.00 | 1.640 | 870.6 |
| 2 | [rs1046724_T_D] => [rs7448421_C_D] | 0.61 | 1.00 | 1.637 | 879.0 |
| 3 | [rs13171869_T_D] => [rs1046724_T_D] | 0.61 | 0.99 | 1.628 | 849.8 |
| 4 | [rs13171869_T_D] => [rs7448421_C_D] | 0.61 | 0.99 | 1.628 | 849.8 |
| 5 | [rs2073950_A_D, rs2301621_A_D] => [rs10849949_C_D] | 0.61 | 0.99 | 1.625 | 858.2 |
| 6 | [rs2832503_G_D] => [rs977779_C_D] | 0.62 | 1.00 | 1.622 | 879.0 |
| 7 | [rs12315146_A_D, rs2301621_A_D] => [rs2073950_A_D] | 0.60 | 1.00 | 1.622 | 826.1 |
| 8 | [rs2301621_A_D] => [rs10849949_C_D] | 0.61 | 0.99 | 1.622 | 854.1 |
| 9 | [rs2073950_A_D] => [rs10849949_C_D] | 0.61 | 0.99 | 1.622 | 854.1 |
| 10 | [rs11682173_T_D] => [rs11692215_T_D] | 0.61 | 0.99 | 1.619 | 845.9 |

#### TABLE 9
#### TOP 10 RULES IDENTIFIED IN CONTROLS

| # | Rule | $\sigma$ | $\delta$ | Lift | $\chi^2$ |
|---|------|----------|----------|------|----------|
| 1 | [rs10501544_C_D] => [rs12280583_T_D] | 0.61 | 1.00 | 1.637 | 1105.5 |
| 2 | [rs10828296_G_D, rs11593316_T_D] => [rs6482203_A_D] | 0.61 | 1.00 | 1.635 | 1089.0 |
| 3 | [rs10828296_G_D, rs1926690_G_D] => [rs6482203_A_D] | 0.60 | 1.00 | 1.635 | 1084.9 |
| 4 | [rs7171993_G_D] => [rs3743121_A_D] | 0.61 | 1.00 | 1.630 | 1113.8 |
| 5 | [rs10828296_G_D] => [rs6482203_A_D] | 0.61 | 1.00 | 1.630 | 1097.1 |
| 6 | [rs11593316_T_D, rs6482203_A_D] => [rs1926690_G_D] | 0.61 | 1.00 | 1.627 | 1080.7 |
| 7 | [rs10828296_G_D, rs11593316_T_D] => [rs1926690_G_D] | 0.60 | 1.00 | 1.627 | 1072.6 |
| 8 | [rs6482203_A_D] => [rs1926690_G_D] | 0.61 | 0.99 | 1.615 | 1059.8 |
| 9 | [rs10828296_G_D] => [rs1926690_G_D] | 0.60 | 0.99 | 1.613 | 1047.6 |
| 10 | [rs2042867_T_D, rs4979935_T_D] => [rs735638_G_D] | 0.60 | 0.99 | 1.604 | 1013.7 |

Several classification tasks are conducted using the top 300, 200, 100 and 50 rules from the ARM analysis, which corresponds to 204, 161, 124, and 92 SNPs respectively. To accomplish this, the SNPs from each set of rules are compressed using SAEs as conducted in the previous experiment (See section 3.3). However, this time by utilising three AEs instead of four (since the number of input features was considerably lower), with a variable number of hidden units. The number of AEs and hidden neurons are arbitrarily selected to gradually reduce the number of initial fea-tures. The final layers of the SAEs are then utilised to initialise the weights of the MLPs before being fine-tuned for classification tasks.

In Fig. 5 the AUC values for the different classifiers are depicted. The different colours in the plot correspond to the different AEs (compression layers) considered in the stack, where the first, second and third layers are represented in blue, orange and green respectively. These results demonstrate that the classifier is not randomly assigning labels to the samples (AUC > 50%).

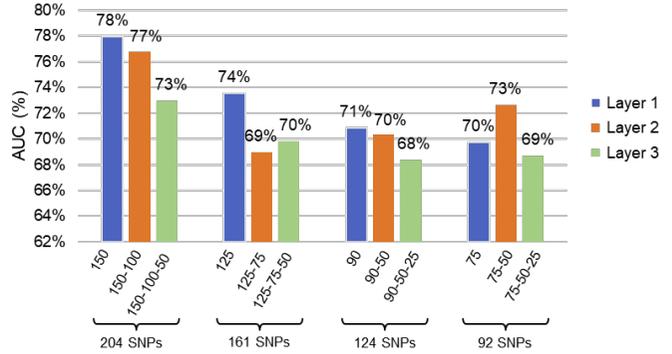

Fig. 5: AUC values for the different classification analyses conducted for the top 300, 200, 100 and 50 rules

Table 10 contains the best classifier performance values for the test set using the SAE and SNPs from the top 300, 200, 100 and 50 rules. In this instance 204 SNPs are compressed using the following layer configuration: 150-100-50. Nevertheless, using two AEs (150-100) achieved the best results. Similarly, 161 SNPs are compressed using a 125-75-50 layer configuration, 124 SNPs are compressed using a 90-50-25 layer fconfiguration, and 92 SNPs compressed using a 75-50-25 configuration.

#### TABLE 10
#### BEST RESULTS FROM SAERMA USING THE TEST SET

| Top Rules | Layers | SE | SP | Gini | LogLoss | AUC | MSE |
|-----------|--------|------|------|------|---------|------|------|
| 300 | 150-100 | 0.77 | 0.68 | 0.53 | 0.58 | 0.77 | 0.20 |
| 200 | 125 | 0.74 | 0.66 | 0.47 | 0.61 | 0.74 | 0.21 |
| 100 | 90 | 0.69 | 0.66 | 0.42 | 0.62 | 0.71 | 0.22 |
| 50 | 75-50 | 0.77 | 0.63 | 0.45 | 0.62 | 0.73 | 0.21 |

Sensitivity, specificity and AUC values presented in Table 10 are depicted in Fig. 6. This represents the best results obtained with SAERMA.

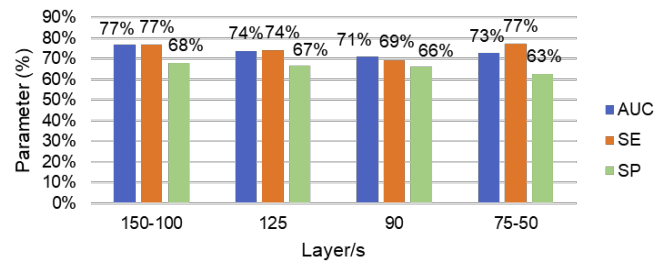

Fig. 6. Best results AUC, SE and SP from SAERMA

## 4 DISCUSSIONS

GWAS can identify common variants with modest to large effects on phenotypes. However, in GWAS studies, SNPs are independently tested for association with phenotypes, without considering the epistatic relationships that exist between genetic variants. Hence, a novel methodology was considered in this study, in which QC and association analysis, performed in GWAS, are combined with ARM and DL stacked autoencoders to detect epistatic interactions between SNPs. A multilayer feedforward artificial neural network classifier is initialised using SNPs and epistatic information learned by the DL SAE (guided and interpreted by ARM), to classify case-control samples from the eMERGE MyCode dataset. The complete network models the epistatic effects of SNP perturbations while ARM provides model interpretation.

In the first experiment, following QC and association analysis, the capacity of the filtered SNPs to discriminate between case and control samples using a GLM was evaluated. Results indicate that GLM can accurately identify case and control individuals using 2,465 features (SNPs) with an AUC of 94% (SE = 85%, SP = 90%, Gini = 87%, Logloss = 0.3288 and MSE = 0.0976) when using the test set, as shown in Table 3. Although AUC values remained high when 248 and 32 SNPs were used as input features (see Table 3), specificities deteriorate when the number of SNPs is reduced. The major limitation with GLM models, however, is that it is not possible to model interactions between SNPs.

In order to address this, the second evaluated experiment modelled MLP NNs. MLPs are non-parametric models capable of capturing complex non-linear relationships between dependent and independent variables through hidden nodes. Using an MLP classifier with the rectifier activation function with dropout regularisation and genetic variants with P-value $< 1 \times 10^{-2}$ (2,465 SNPs) it was possible to obtain SE = 95%, SP = 98%, Gini = 99%, LogLoss = 0.1061, AUC = 99% and MSE = 0.0291. In contrast, using 5 SNPs (P-value $< 1 \times 10^{-5}$) resulted in a significant performance drop (SE = 99%, SP = 0.6%, Gini = 21%, LogLoss = 0.6750, AUC = 60% and MSE = 0.2410). The results indicate that the model was unable to correctly recognise actual negative cases (i.e. non-obese individuals).

Acceptable results were obtained using MLPs with 2,465 and 248 SNPs, with high AUCs and relatively balanced SE and SP values as shown in Table 5. However, compared with the GLM experiment, specificities deteriorate when the number of input features reached 32 or less. These results reveal that MLPs achieve overall better results than GLM, probably due to the nonlinear nature of the interactions occurring between SNPs.

While the MLP can learn and capture epistatic information, a high number of features are required to achieve good performance. It is not clear to what extend those SNPs interact and what proportion of the data actually represents noise. Investigating this further, autoencoders were used to determine if a low-dimensional representation of our input data (2,465 SNPs) could be achieved, while retaining all relevant information. This helps to remove any redundant features with a particular focus on epistasis.

Therefore, in the third exdperiment, a set of 2,465 SNPs

(P-value $< 1 \times 10^{-2}$) and four single layer AEs were implemented to compress SNPs through 2,000-1,000-500-50 hidden units. The best result with the test set was obtained using 2,000 hidden units (SE = 95%, SP = 93%, Gini = 95%, Logloss = 0.1956, AUC = 97% and MSE = 0.054057) and a rectifier activation function with dropout regularisation. Conversely, the worst result was achieved when the features were compressed to 50 hidden units (SE = 78%, SP = 80%, Gini = 70%, Logloss = 0.476864, AUC = 85% and MSE = 0.156315), which are still encouraging. See Table 7 for details.

Although a gradual deterioration in performance is observed, the classifier performance is still high even with 50 units, over 85% AUC with SE = 78% and SP = 80% with no evidence of overfitting. This supports our previous argument and shows that there is significant noise within the initial 2,465 SNPs. This, thus, demonstrates the potential of the proposed deep learning methodology to abstract large, complex and unstructured data into latent representations capable of capturing the epistatic effect between SNPs in GWAS.

Sensitivities and specificities are generally more balanced – for example, compare the results in Table 5, for 32 SNPs (P-value $< 1 \times 10^{-4}$), where SE = 95% and SP = 29% with those in Table 7, for 2,000-1,000-500-50, where SE = 79% and SP = 80%. In addition to SE, SP and AUC values, SAEs also improved Gini, Logloss and MSE values when compared with models using a similar number of input features. More importantly, the results obtained using SAEs with 50 hidden nodes are close to those achieved with 248 SNPs using the GLM and MLP. A summary of these results is shown in Table 11.

TABLE 11
RESULT COMPARISON FOR GLM, MLP AND SAE USING 248, 248 AND 50 FEATURES RESPECTIVELY IN TEST SET

| Model | Feat. | SE | SP | Gini | LogLoss | AUC | MSE |
|-------|-------|------|------|------|---------|------|------|
| GLM | 248 | 0.95 | 0.63 | 0.78 | 0.41 | 0.89 | 0.13 |
| MLP | 248 | 0.90 | 0.79 | 0.85 | 0.35 | 0.93 | 0.11 |
| SAE | 50 | 0.78 | 0.80 | 0.70 | 0.48 | 0.85 | 0.16 |

The SAE experiment provides a novel approach for feature extraction and classification tasks, using latent information extracted from high-dimensional genomic data. This allows us to screen individuals with higher predisposition to obesity. However, compressing the features using SAEs alone makes it difficult to identify which of the 2,465 SNPs contributes to the compressed hidden units. This is a well-known problem in neural networks where model interpretation is difficult to achieve. In order to address this issue, the final experiment combines the strength of SAE and ARM via the Apriori algorithm, to provide an interpretation of the DL networks utilised in this study.

ARM is more transparent than other machine learning algorithms as it provides knowledge based explanative rules, serving therefore as a white-box model. Hence, this approach allows us to investigate relevant epistatic patterns and determine the direction of associations between SNPs, while SAE and MLP classification provides an objective performance measure to validate the models ARM produces. These are tightly correlated in that altering the interest measures (support and confidence) in ARM impacts on the performance metrics of the SAE and MLP models.

In the rule generation process, redundant rules are removed to alleviate the high number of rules being generated in the rule mining which aids computational efficiency. Although lift values for all the top 10 rules in cases and controls were slightly higher than 1, the dependency of the rules was supported by very high values of $\chi^2$. The inference made by an association rule does not necessarily imply causality. Counterwise, it suggests a strong relationship between SNPs in the antecedent and consequent of the rule. Hence, ARM results need to be carefully interpreted.

The rules generated help to reveal new insights in obesity as a complex disease. While the genes in rules 1, 2, 3, and 4 in cases have not been associated with obesity, the genes in rules 5 to 10 reveal something different. In fact, the ATXN2 gene present in rules 5, 7, and 9 has been involved in severe early onset obesity in children [37]. In rule 7, the ATXN2 gene also interacts with the MAPKAPK5 gene, which has shown gender-dependent differences in anxiety-related processes and locomotor activity in mice [38]. A weak but positive association between anxiety and obesity in humans has been reported, although further studies were recommended in [39]. Furthermore, the GRIK1 gene in rule 6, has been reported as a novel obesity candidate gene that may contribute to highly penetrant forms of familial obesity [40]. Finally, the gene AFF3 in rule 10, has shown associations with triglycerides in Asian populations [41].

One of the possible reasons why obesity related variants within the genes FTO or MC4R were not identified in any stages of the proposed methodology may be due to the effect of removing a very large number of variants by using stringent thresholds in the per-marker QC step. It is known that statistical power to detect a SNP of a given effect via GWAS increases with both sample size and the density of genetic variants across the genome. In this study, the sample size is relatively small (1,997 individuals after QC) and the density of markers was also reduced considerably (240,950 SNPs after QC). In the top 10 rules identified in cases, the genetic variants identified within or close to the SGOL2, AOX1, ZNF354B, ZFP2, ATXN2, MAPKAPK5, GRIK1 and AFF3 genes form interactions. Four of these genes have implications with obesity related traits (ATXN2, MAPKAPK5, GRIK1 and AFF3) whereas SGOL2, AOX1, ZNF354B, ZFP2 have not previously been associated with obesity. In Table 12, the SNPs within the top 10 rules in cases have been summarised. The table includes five columns with information about the SNP ID, risk allele, overlapped or closest gene, whether it has been previously associated with obesity related traits or not, and the name of the related trait in cases of previous association.

After rule mining was applied to the filtered SNPs (2,465 SNPs), several classifiers were pre-trained with the compressed units extracted from the top 300, 200, 100 and 50 rules. For each set of rules, their SNPs (forming the rules) were used as input features for several SAEs. Then, the MLP classifiers were initialised and fine-tuned with the final hidden layer of the SAE. The best results are presented in Table 10.

TABLE 12
SUMMARY OF GENETIC VARIANTS IDENTIFIED USING ARM IN CASES OF OBESITY

| SNP ID | Allele | Gene | Previous Assoc. | Related trait |
|---|---|---|---|---|
| rs12053340 | C | SGOL2 | No | - |
| rs1527944 | T | AOX1 | No | - |
| rs1046724 | T | ZNF354B | No | - |
| rs7448421 | C | ZNF354B | No | - |
| rs13171869 | T | ZFP2 | No | - |
| rs2073950 | A | ATXN2 | Yes | Severe early onset obesity |
| rs2301621 | A | ATXN2 | Yes | Severe early onset obesity |
| rs10849949 | C | ATXN2 | Yes | Severe early onset obesity |
| rs2832503 | G | GRIK1 | Yes | Familial obesity |
| rs977779 | C | GRIK1 | Yes | Familial obesity |
| rs12315146 | A | MAPKAPK5 | Yes | Anxiety-related processes and locomotor activities |
| rs11682173 | T | AFF3 | Yes | Triglycerides |
| rs11692215 | T | AFF3 | Yes | Triglycerides |

In the first set of 300 rules with 204 SNPs, the best result in the test set was achieved when the input features were compressed to 100 units, with an AUC of 77%, SE = 77%, SP = 68%, Gini = 53%, Logloss = 0.5769 and MSE = 0.1968, as shown in Table 10. Although a higher AUC was achieved with a single AE and 150 hidden units (AUC = 78%, SE = 80%, SP = 63%, Gini = 56%, Logloss = 0.5770 and MSE = 0.1952), the sensitivity value was inferior. In these situations, it is up to the expert/clinician to decide whether it is more important to detect cases of obesity more accurately than normal individuals. However, in this study, the capacity of our proposed solution to detect cases and controls in a balanced manner has been prioritised. This means that results with a balanced SE and SP and high AUC were selected.

Using the top 200 rules (161 SNPs) and the above criteria, the best result in the test set was accomplished when a single AE and 125 hidden units were used as input for the MLP classifier (see Table 10). The classifier achieved an AUC = 73% with SE = 74% and SP = 66% (Gini = 47%, Logloss = 0.6099 and MSE = 0.2104).

Using the top 100 rules with 124 SNPs, it was possible to achieve 71% AUC with 69% sensitivity and 66% specificity (Gini = 42%, Logloss = 0.6231 and MSE = 0.2167) by compressing 124 SNPs down to 90 units.

Finally, the models trained with the lowest number of features (92 SNPs from the top 50 rules) achieved the best classification results using a 75-50 layer configuration (See Table 10). This model reached an AUC value of 73% with sensitivity and specificity values of 77% and 63% respectively (Gini = 45%, Logloss = 0.6178 and MSE = 0.2142).

Even though the best results were achieved in the largest set of SNPs (300 rules), we observed that some of the models were able to compress the features down to 50 hidden units and get over 70% AUC, as can be seen in Fig. 5. Additionally, the AUC, SE and SP values from the best models achieved by SAERMA are depicted in Fig. 6. The results indicate that there is not much variation between the performance values (AUC, SE and SP) among classifiers despite the reduction in the number of SNPs and hidden units within the AEs.

Therefore, the best overall result from the different classifiers was AUC = 77%, attained by 100 compressed units from the top 300 rules as can be observed in Table 10. The classifier was able to classify obese individuals (SE = 77%) more effectively than normal samples (SP = 68%). These results can be achieved with a maximum of 204 SNPs although the SAE is able to reduce noise and achieve that value (AUC = 77%) with 100 hidden neurons (this is a 50.99% reduction in the feature space). However, it is not possible to accurately determine which of those 204 SNPs correspond to the 100 compressed hidden neurons.

For a more granular mapping of the interactions between SNPs, we can refer to the top 50 rules result (92 SNPs), where the input was compressed to 50 hidden units (see Table 10). Even though dimensionality reduction in this case affects the performance of the classifier with respect to the best result (using 204 SNPs), the SE value remains the same (77%), while SP is reduced by 0.05% and AUC by 0.04 %. Thus, it is true to say that the 50 hidden nodes representing epistatic interactions can be interpreted using the 92 SNPs selected by ARM. Although this does not represent a full interpretation of the results obtained using SAEs, the approach presented in this paper provides a close approximation of the epistatic interactions that likely occur in the MyCode data.

The best overall performance was achieved by the SAE using 204 SNPs. Hence, utilising SNPnexus [42] it was possible to query the 204 SNPs and report the overlapped or closest genes according to the GRCh37 assembly. A table containing genomic annotations for the 204 SNPs reported in this study has been included in the supplemental material (Table S1 in File 2). It is expected that these findings will help future researchers to better understand how epistasis in obesity occurs using genome-wide data, providing candidate SNPs to investigate obesity further.

## 5 CONCLUSIONS AND FUTURE WORK

Overall, the results in this study highlight the benefits of using deep learning stacked autoencoders to detect epistatic interactions between SNPs in genomic data and how these can be used to model MLPs to classify obese and non-obese observations from the eMERGE MyCode dataset. This contributes to the computational biology and bioinformatics field and provides new insights into the use of deep learning algorithms when analysing GWAS that warrants further investigation. However, the minute non-linear transformations of the input space that occur in the autoencoders, makes it very difficult to trace the amount of variance they contribute from case-control data. This is a common problem in neural network modelling that seriously hinders genomic analysis. To aid with this issue, association rule mining was used in combination with stacked autoencoders. This allowed us to identify patterns in the form of rules which represent interactions between a filtered subset of SNPs. The benefits of incorporating rule mining to the proposed pipeline were twofold. First, it allowed us to generate significant rules and plot their interactions. Second, feeding the stacked autoencoders with the most significant rules allowed us to obtain dynamic classification performances by adjusting the number of rules generated in the rule mining process, serving thus as a validation and interpretation technique for epistatic feature extraction in the neural network utilised in the study. Adjusting support and confidence coefficients to increase the number of rules also requires more computational complexity. Therefore, in this study only rules generated with support and confidence values of 0.6 and 0.8 respectively were presented. This allowed us to empirically produce the best results without reaching computational overload with the resources available.

While work exists in biological analysis of variants that alter functional regulatory elements (i.e. elements that control gene expression and DNA) using deep learning methods [15] and epistasis analysis based on frequent itemset mining using the Apriori algorithm [13], to the best of our knowledge this research is the first comprehensive study of its kind that combines GWAS quality control and logistic regression with association rule mining and deep learning stacked autoencoders for epistatic-drive GWAS analysis and case-control classification.

Several novel contributions have been provided using the proposed methodology. However, there are still areas for improvement. In future work, biological validation of the rules identified by SAERMA needs to be provided. A common approach to achieve this is via gene set enrichment analysis which is based on the functional annotation of gene sets. Any identified rules including more than one gene involved in a particular pathway can be considered potential true obesity epistasis.